# Fast tree numeration in networks with synchronized time

*Khuziev Ilnur[1]*

**Abstract.** In this article we present a protocol for building dense numeration in network with unknown topology. Additionally to a unique number each node as result of the protocol will get information about a spanning tree. This spanning tree is constructed during BFS search from the leader node. This property of the numeration can be useful in other tasks, as example we present a protocol for searching bridges in network. The time of numeration building in our protocol is linear in network size, simple informational lower bounds also linear (it is required at least linear number of bits for code tree structure). In bridges searching problem our protocol also heats lower linear bound: in result each node knows about all bridges.

## *Introduction*

Distributed computations are hardly investigated last time, some basics can be found in [1], [2]. The main concept is *a network* – a set of nodes and bound between them, nodes a participants of an interactive protocol, bounds are used for information transmitting. The range of problems and available methods depends on used network model, and there are many different models. Therefore, in the first chapter we focus on formal definition of used in this work model. The main points of used model are following:

1. synchronous time – transition through all bounds happen in the same moment;
2. limited bandwidth of bounds – during one communication round only constant number of bits can be transmitted through each bound;
3. lack of information about network topology – protocols we study must be correct in some family of networks, at the beginning of protocol execution nodes do not known in which exactly network from considered family they are;
4. *uniformity* of protocols – each node execute the same program, which must solve a considered problem in each network from family.

The same model of network are investigated in, for example, in works [3], [4], but in more specialized way, we will use definitions that are more generic. Other popular models with synchronized time use either full information about topology either prior dense numeration, or both. Also, popular property for networks models is "signed" messages: bandwidth of bounds is $O(\log V)$, there $V$ is number of nodes, so each message can contain the number of the author.

Mainly we use family of networks with a leader: exactly one node at the beginning of execution knows that it is leader, and other nodes know, that we are not. This model can be considered as basic, because networks with unique numbers (they form sparse numeration) is reducible to the model with a leader [4], meanwhile in networks without unique identifiers most problems are not solvable [4].

**The main result of the work** (section 2, Theorem 1) is protocol for building dense numeration (each node will get unique number from segment$[1, V]$) in network with a leader, the protocol finishes in linear time. In other words, we reduce network with a leader into more researched model. Note, that linear time of work good enough, because protocol in distribute information about a spanning tree: each node will know structure of some tree and its place inside this tree. Therefore, each node will receive sufficient part of information about network's topology. Note that building spanning tree is a basic task for solving many problems. Additionally, as information about tree is linear, we get linear lower bound (a node with constrained degree cannot receive more than constant bits of information per round).

To sum up, the model we use from one side is more basic, from other side it allow to solve some problems more effective. For example in work [6] presented a protocol for searching bridges in model with signed messages, which solves the problems in linear time. If we reinterpret that protocol in basic model, the time bound will be $O(V\log V)$, that worse than linear bound for protocol we build in section 3. In addition the protocol we build is better by total traffic through network.

Note, that definition of *time of protocol execution* is not trivial: it is required either synchronized stop of all nodes either lack of conflicts if some nodes already stopped. Formal definitions we .give in section 3.

Main stages of protocol we build in section 3 is following:

1. Finding out "height" of network – maximal distance between leader and other node. This step partly technical and required for synchronize starting each next stage

---

[1] Khuziev Ilnur, MIPT, Ilnur.khuziev@yandex.ru

2. Tree numeration (protocol described in section 2). In result each node get unique number from segment $[1, V]$ (there $V$ – number of nodes) and also it get structure of some network's spanning tree, more other number it got is number of the node in this tree.
3. Nodes send their number to each neighbor. As result, at first each edge get number (as pair of numbers of adjacent nodes); at second nodes find out place of neighbors in the tree build in second stage.
4. Nodes transmit to tree-parents information based on previous stage. As result each node find out adjacent bridges.
5. Last optional stage allow share information between all nodes in linear time.

Note that idea of this protocol is very similar to protocol from [6], but using other model and evicting unnecessary data from messages allow to improve protocol time. I would like to emphasize that the Pritchard's protocol reaches lower bound but only in model with signed messages, reinterpreting the protocol in more generic model do not save optimality.

## *1. Definitions*

For make definition of protocol easier, we define network with help of *index function*.

**Определение.** The tuple $G = (V, \deg, E)$ is called network, there the first element – the set of nodes $V$;

the second element – *degree function* $deg: V \to \mathbb{N}$;

the third element – index function $E: V \times \mathbb{N} \to V \times \mathbb{N}$, which produce bounds in the network

Index function must satisfy following conditions:

1. $E(v, k)$ defined only for $k \in [0, \deg(v)]$
2. $E(E(v, k)) = (v, k)$
3. $E(v, k) \neq (v, k)$

Note, in this definition we do not prohibit loops and multiple edges. In problems we investigate, they do not play any significant role. For some simplification, we consider only connected graphs without loops and multiple edges.

We also define functions $p_E(v, k), r_E(v, k)$, by equation

$$E(p_E(v, k), r_E(v, k)) = (v, k)$$

Further we define *a protocol in network,* rules of updating nodes' states in network with given initial state. Our goal is to give definitions, there available to each node information defines explicitly. To unite information in node and topology of network we define a *configuration* – a pair of network and nodes' initial state. It allows us to look on a protocol as a transformation of configurations: initial state and information got by a node during step can be taken as a new initial state.

**Definition.** We call arbitrary function $I: V \to 2^{\mathbb{N}}$ as *initialization function of nodes.*

In fact, we do not set any constraints on $I$ and allow to use any countable structure. Although in all examples in this work we use only finite states, countable sets useful in definition of probabilistic protocols (initial state can contain infinite binary string).

*Configuration is a fourth* $(V, deg, E, I) = (G, I)$, determines both topology of network and initial states of nodes.

*Family of configurations with leader* is a set of configurations $(G, I)$, there $G = (V, deg, E)$ can be any finite connected network (also, as we mentioned before, without loops and multiple edges), and function $I: V \to \{0,1\}$ is equal to one on exactly one vertex from $V$.

In this article we work with this family. Reducibility for this family in meaningful situation is described in [4] (in anonymous networks due to symmetry is impossible to find out topology).

Note, that the leader mainly used for synchronize actions of other nodes. The solution of form "transmit all information to the leader, get answer from him" is valid, but at first this solution is not trivial (it is needed to make numeration of nodes at first) and, at second, that kind of solutions can be not optimal (in particular, in problem of bridges searching).

Let fix an alphabet $\Sigma$ – a set of symbols, that can be transmitted through bounds. For convenience let $\Sigma$ contains at least 3 elements and one of them is $\lambda \in \Sigma$, that we call "empty symbol". In description of protocols, if no rule specify other behavior, node sends $\lambda$ through bound.

**Definition.** *Protocol* is an arbitrary function with arity 5

$$\pi(d, t, s, i, k) \in \Sigma,$$

there $d, t, k \in \mathbb{N}, k \leq d, s \in \Sigma^{t \cdot d}, i \in 2^{\mathbb{N}}$. In other words, protocol $\pi$ – is a function, which by degree of vertex $d$, its initial state $i$, previous history of communication $s$, moment of time $t$ determines a symbol to be transmitted through k-th bound of node.

**Definition.** *State function* $H$ of network $G$ – is the function that satisfy

$$H(t, v) \in \Sigma^{t \cdot \deg(v)} \text{ and } H(t, v) = (H(t-1, v), H'(t, v)).$$

We associate $H(t, v)$ with a set of messages that vertex $v$ got till moment of time $t$. The second condition formalizes that $H(t, v)$ should satisfy consistency requirement: $H(t, v)$ must be done from $H(t-1, v)$ by adding $\deg(v)$ symbols into history. Through $H'(t, v)$ we denote a set of incoming messages of $v$ at round $t$.

With $H'(t, v, k)$ we denote $k$-th symbol of $H'(t, v)$, in other words, $H'(t, v, k)$ is the symbol that vertex $v$ got through by k-th bound.

Now we ready to define how protocol $\pi$ recursively defines state function $H_\pi$ by configuration $(V, deg, E, I)$:

$$H'_\pi\big(t, p_e(v, k), r_e(v, k)\big) = \pi(\deg(v), t-1, H_\pi(v, t-1), I(v), k)$$

Note, that the definition is correct, because pairs of form $(p_e(v, k), r_e(v, k))$ and pairs of form $(v, k)$ consist a matching, in other words while pairs $(v, k)$ go through all valid values, the same do pairs $(p_e(v, k), r_e(v, k))$, therefore $H'_\pi$ is defined for all $(v, k)$, for induction base we take база $H_\pi(0, v) = \emptyset$.

Now let move to the problem of protocol stopping. We want to work with protocols that have finite execution time. There are many variants to define finiteness of execution time (see [1] and [4]), in this work we mainly work with *strong-processor-terminating* protocols – informally, it is such protocols, that have finite execution time and all noes finish they work at the same moment.

**Definition.** We call protocol $\pi$ strong-processor-terminating in family of configurations $\{(G, I) \mid (G, I) \in A\}$ if exists a function $end: (d, i, s, t) \mapsto x \in \{0,1\}$, such that for any configuration $(G, I)$ from the family there exists moment of time t, such that

1. $end(\deg(v), I(v), H_\pi(v, t'), t') = [t' \geq t]$, here $[...]$ – is indicator function;
2. state $s = H_\pi(v, t)$ is *terminating* in protocol $\pi$ – satisfy condition
$$\pi(\deg(v), t', ss', I(v), k) = \lambda, \forall t' \geq t, \forall k, \forall s' \in \Sigma^{\deg(v)(t'-t)},$$
this condition formalizes requirement "for all future input messages node will note send nothing except empty message».

The minimal t, such that the condition 1 holds, called the execution time of protocol $\pi$ in configuration $(G, I)$.

We want make accent that end-indication-function (named end in previous definition) is the same for all configurations from the family. This function allow locally (only by information available in node) determine the moment to stop execution.

We almost have done with all definitions, the last question to discuss is "where to get the answer". So let's define the problem and solution. Our task is to compute some function $f(G, I, v)$, that can depend on network topology, initial states and vertex number. We say that we solve task $f$, if after execution stopping in all nodes of network the value of $f(G, I, v)$ is computable[2]. More formally: there exists $solve$, that

$$\forall (G, I) \in A \ \forall v \in G \ solve\big(H_\pi(t, v)\big) = f(G, I, v), \text{ where t – execution time of protocol } \pi \text{ in configuration} (G, I).$$

We want to make an accent on the fact, that function $solve$ must be the same for all configurations.

---

[2] Note that we do not mean Turing-computable in this context (and indeed in whole work too), computable means mapping existence

For example, in bridges searching problem we want each node to be able compute the list of adjacent bridges.

Also we will investigate harder problem: it is required for all nodes to know the list of all bridges in network. This formulation is quietly invalid, because in family with a leader there aren't any fixed numeration of edges. This difficulty can be solved by additional requirements: at first, each node must get an unique number, at second each edge numerated as a pair of numbers of adjacent vertexes. See theorem 3.

## Protocol of searching height

In this subsection we describe simple protocol, that we will use further for make compositions of protocols (see the next subsection).

Let we have a network with a leader, we donate the leader as L. As alphabet we take $\Sigma = \{0,1,2,\lambda\}$. Let $D$ denote the maximal distance максимальное расстояние between L and other vertexes. In other words, D is the height of the BFS-tree with L as the root.

**Statement**. There is strong-processor-terminating protocol with execution time $3D + 4$, such that in result each node can compute $D$.

The rules of nodes actions, that define the protocol are following:

1. L send to all its neighbors message «1».
2. Each node that got message «1» first time, also send all neighbors message «1» (this rule have exceptions, see point 3). The bound, which send to the current node the first message «1», is called parent bound. Note, that there can be more than one such bounds (they have send message «1» in the same round), in such case the choice of parent bound is arbitrary. The moment then parent bound is formed in node $v$ is called $P_t(v)$. Also we define $P_t(L) = 0$.
3. Each node as the answer for all further messages «1» (and for messages that have been received in the round $P_t(v)$ but by other – not parent – bounds) send message «0».
4. If a vertex from all neighbors, except parent node, got at least once message «0» it send to parent node message «0».
5. Then L got «0» from all neighbors, it send all neighbors message «2».
6. Each node that get message «2», also send «2» for all child nodes. The moment of sending «2» by vertex $v$ to its child is denoted $S_t(v)$
7. Each node that executed step 6 (or 5 for leader) stops after $\frac{(S_t(v) - P_t(v) - 4)}{2} - P_t(v)$ rounds.

The proof that protocol stops at round number 2D, so the ability to compute D will be corollary.

Easy to see that in this protocol we just make simple broad-first-search. At each round distance between L and nodes that received «1» increased by one. In other words $P_t(v)$ is equal to distance between $v$ and L. If node $v$ is a *leaf* (all neighbors of $v$ have the same or smaller distance to L)and distance between $v$ and L is equal to k, then at round k+1 this node send «1» to all neighbors (except parent), on round number k+2 node $v$ will receive «0» as answer. Therefore leaf node $v$ at round k+3 have to send «0» to parent. Then by induction we get statement that at round 2D+3 node L will get «0» from all neighbors. Similarly we prove equation

$$S_t(v) = 2D + 4 + P_t(v).$$

Therefor the moment of stopping (by point 7) of node $v$ is

$$S_t(v) + \frac{(S_t(v) - P_t(v) - 4)}{2} - P_t(v) = 3D + 4.$$

Note, first 4 rules from list above form rules for *echo-protocol*.

## Clock synchronization method

It is easier to describe process-terminating protocols (do not require stop execution at the same time in all nodes). In this subsection we show how in networks with a leader from arbitrary process-terminating protocol make strong-process-terminating protocol. The overhead of this transformation is O(D).

Let protocol $\pi$ use alphabet $\Sigma$ and all nodes in finite stop execution (note that by definition the moment of stopping is computable in nodes). We will build a new protocol $\pi'$, that will work over wider alphabet, for example $\Sigma' = \Sigma \times \Sigma$. This technic allow us to execute two protocols at the same time: at each round pair of symbols are send through bounds. The first element of pair i s associated with messages of protocol $\pi$, the second message is associated with an echo-protocol.

At the beginning of $\pi'$ it differ from $\pi$ only by send message: instead message $\sigma \in \Sigma$ it sends message $(\sigma, \lambda)$. Then by the rules of $\pi$ the leader switches to stop-state, he starts echo-protocol (by sending corresponding messages in second component of message, first component stay empty). Each node $v$ execute steps of echo-protocol (by sending corresponding messages in second component of pair) only after switching to stop-step by rules of $\pi$. If some node have not finished execution of $\pi$, but got not empty symbol in second component, the node memorize fact of receiving that message and it answer on that message after switching to stop-state in $\pi$.

Hence, the echo-protocol finishes (the leader got answer from all nodesлидер получит ответы от всех дочерних узлов) only after all nodes have finished execution of protocol $\pi$. After finishing echo-protocol nodes start to execute protocol of searching height also using second component of messages. It is crucial that protocol of searching starts to execute by leader and it stay correct if nodes have started after arbitrary delay.

If during protocol of searching execution node switches to stop-state, the node switches to stop state in $\pi'$ also. As protocol of searching is strong-processor-terminating, all nodes will switch to the stop-state at the same time, therefore $\pi'$ is strong-processor-terminating.

To sum up, we have shown how to transform any processor-terminating protocol into strong-processor-terminating protocol with time-overhead $O(diam(G))$. This allow us to easy make compositions of processor-terminating protocols with small overhead (in fact at almost all meaningful situations this overhead does not increase asymptotic), the fact is not trivial because basic-protocols can require start of execution at the same time in all nodes.

## 2. Tree numeration protocol

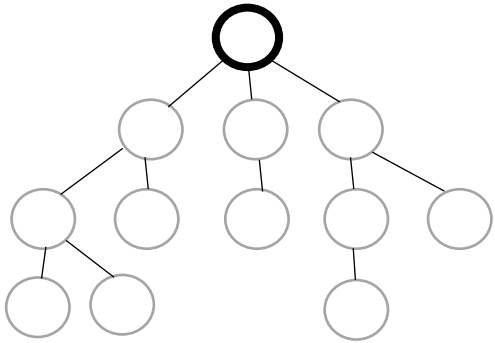

Image 1 Tree with root

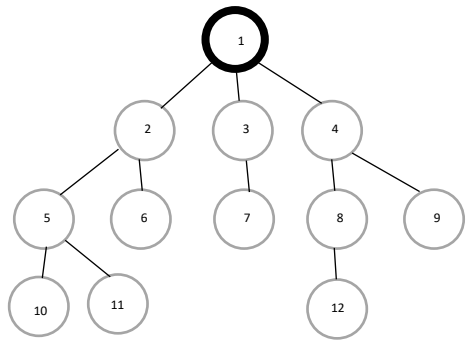

Image 2 Numeration via BFS

Let we have a tree with a root. Then we automatically get partitioning of nodes on slices (by distance to the root), also we have well-defined relation parent-child. Let fix ordering of child-nodes at each node. Then we have define ordering on all vertexes by following rules:

1. If nodes $v$ and $u$ have different distance to the root, then $v < u$ iff distance from $v$ to the root is less than distance from $u$ to the root
2. If $v$ and $u$ have the same distance to the root and they are child nodes for the same node $z$, then ordering between $v$ and $u$ is defined by ordering on child-nodes of $z$.
3. If nodes $v$ and $u$ have the same distance to the root and they have not common parent, then ordering of $v$ and $u$ is the same as ordering of their parents (defined by induction).

The numeration induced by this ordering we name **numeration via BFS**.

**Definition.** As *BFS-certificate* of tree $T$ we call a binary string with $|V|$ blocks, the i-th block consist of k symbols "1" and one symbol "0", there $k$ – number of child nodes of vertex with number $i$ (number of vertex is defined by numeration via BFS).

*Example*. For tree from image 2 the certificate is following: $1110 - 110\ 10\ 110 - 110\ 0\ 0\ 10\ 0 - 000$ (dashes divide parts corresponding for different slices).

**Definition.** As *certificate of node i* in tree $T$ we call a string that differs from *BFS-certificate* at exactly one symbol: $(i-1)$-th symbol "1" is replaced by symbol "2".

*Example*: vertex number 7 from image 2 have following certificate: $1110 - 110\ 20\ 110 - 110\ 0\ 0\ 10\ 0 - 000$

By vertex's certificate, it is possible to restore following information:

- number of vertex
- number of parent vertex
- path in tree to any other vertex (if it number is known)

In other words, vertex's certificate give full information about both tree structure and place of the vertex in that tree.

Let denote BFS-certificate as $S(T)$. Through $T_v$ we denote subtree of vertex <u>v.</u>

*Example*: for tree from the image 2 holds $S(T_2) = $ 110 110 0 00.

**Statement 2.1.** BFS-certificate is prefix-computable by certificates of subtrees of child-nodes. In other words if in tree $T$ the root have $k$ child nodes, strings $S_2, S_3, \ldots, S_{k+1}$ are *l*-prefixes of strings $S(T_2), \ldots, S(T_{k+1})$. The *l+1* prefix of string S(T) is computable by $S_2, S_3, \ldots, S_{k+1}$.

**Proof.** Indeed, first $k + 1 > 0$ symbols of string S(T) can be computed with knowledge 0-prefixes of child certificates. The part of certificate corresponding to *m*-th layer is concatenation of parts of $S(T_2), S(T_3), \ldots, S(T_{k+1})$, that correspond to (*m-1*)-th layers of them. Therefore at least l symbols of certificate corresponding to layers 2 or more can be written if known $S_2, S_3, \ldots, S_{k+1}$.

We denote as $S_v(T)$ the certificate of vertex *v* in tree $T$.

**Statement 2.2.** Let vertex *v* have child vertexes $v_1, \ldots, v_k$ in tree $T$, then *l*-prefixes of strings $S_{v_1}(T), \ldots, S_{v_k}(T)$ are computable by known *l*-prefix $S_v(T)$.

**Proof.** Indeed, if *l*-prefix $S_v(T)$ have no symbols "2" then at each of $S_{v_1}(T), \ldots, S_{v_k}(T)$ there is also no any symbols "2" in positions if numbers less or equal to *l* (numbers of child nodes by construction of ordering is greater than number of parent). Therefore, *l*-prefixes of $S_{v_1}(T), \ldots, S_{v_k}(T)$ equal to *l*-prefix of $S_v(T)$. If symbol "2" exists in $S_v(T)$ then the number of *v* is computable number, let denote this number as *n*. Hence it is possible to find out borders of block number *n* (for all of $S_{v_1}(T), \ldots, S_{v_k}(T)$ symbol "2" stay inside n-th block in positions $1, \ldots, k$ correspondingly), so it is possible to place "2" to right place if it must be *l*-prefixes of $S_{v_1}(T), \ldots, S_{v_k}(T)$.

Now we ready to formulate and proof the main result of the work.

**Theorem 1.** There is strong-processor-terminating protocol in network with a leader, such that in result of work each node *v* can compute $S_v(T)$ for some spanning tree $T$, time of execution is O(V).

Protocol consist of two stages: the first one build tree with leader as root (for example, BFS-tree), the second stage build and distribute node certificates.

Let concentrate on the second stage (first one described in protocol of height searching, composition also already explained).

At the same time there is two streams of data: first one from leaves to root build certificates of subtree and send that information to the root; the second stream of data distribute global certificate from root downstream to leaves, during to this distribution nodes send certificate to child with appropriate fixing certificate into child-node certificate.

In accordance with statement 2.1 each node *v* can compute prefix of BFS-certificate corresponding to subtree $T_v$ and send each bit to parent. As child send their certificates, length of known prefix of $S(T_v)$ grows by 1 each round (until whole certificate is not known).

The root in correspondence with statement 2.1, compute prefixes of $S(T)$ and send to child vertex-certificates in correspondence with 2.2. Each non root vertex also with statement 2.2 can compute prefixes of child-nodes and send them.

By construction, it is possible by certificate prefix if it is full (number of zeroes greater than number of ones by one). Therefore each node can stop execution after sending to child-nodes the last symbol of certificate.

Bound on time is abvious: there is delay $O(D)$ before leaves get first symbol of certificate. Length of certificate is linear $O(V)$, additional $O(D)$ we get because of clock-synchronizations.

### *3. Bridges searching protocol*

As example of usage tree-numeration protocol we present protocol for bridges searching in network with a leader.

**Theorem 2.** *In networks with a leader the task of bridges searching is solved by strong-processor-terminating protocol in linear time.*

As we shown above network with a leader is reducible to a network with tree-numeration in linear time. Therefore, it is sufficient to solve the problem in network with numerated tree.

The first fact: all bridges are edges of the tree. Hence we need to find out which of bonds between parents and children are bridges. Our idea is to reduce the task to computing some prefix-computable function. Then we as in previous theorem use distribution mechanism to compute results.

We will use following notations:

- $h_T(v)$ is height of vertex $v$ – distance by tree-edges of $T$ from $v$ to root. As we use only one tree, we skip subscript further.
- For any cross-edge $(v_1, v_2)$ (edge that not part of tree $T$) we denote $P(v_1, v_2)$ as minimum of $h_T(v_1)$ and $h_T(v\_2)$.
- $N(v)$ is a set of all neighbors of $v$, except neighbors of $v$ in tree $T$.
- Minimum of $P(v, v_2)$ there $v_2$ goes through $N(v)$ – is minimal height of cross-edges of vertex $v$, we denote this value as $P(v)$. For vertexes with empty $N(v)$ we define $P(v) = h(v)$.
- The set of children of $v$ in the tree is denoted as $C(v)$.
- Recursively defined function $g(v) = \min(\min_{u \in C(v)} g(u), P(v))$ – is minimal height of cross-edges, that goes from subtree of $v$. For leave vertexes we set $g(v) = P(v)$.

**Statement 3.1** *Edge between vertex $v$ and its parent is bridge iff $h(v) = g(v)$.*

The proof is simple and we leave it as a simple exercise.

So, computing $g(v)$ is enough to find out if edge to parent is bridge, but $g(v)$ is not prefix-computable by information from child-nodes. We will work with function $g'(v)$:

$$g'(v) = H - g(v) = \min(\min_{u \in C(v)} g'(u), H - P(v)).$$

Let $s'(v)$ denote unary representation of value $g'(v)$ with additional zero as last symbol (zero symbol allow to determine if prefix of $s'(v)$ is full or not). It is easy to see that (due to propertis of min) k-prefix of $s'(v)$ is computable with known k-prefixes of $s'(u)$ for all $u \in C(v)$ and known $P(v)$.

The scheme of result protocol is following:

1. Build tree-numeration and distribute BFS-certificate
2. During $\log V$ rounds all nodes send by all bounds their number in tree made at first stage. This allow to all vertexes know numbers of all neighbors. Therefore with help of certificate value of $P(v)$ is computable in all nodes.
3. Each node $v$ computes prefixes of $s'(v)$ with help of information from children (leav nodes know $s'(v)$ exactly after finish of 2nd stage) and send bits to the parent.
4. After getting whole information from children each node knows if bound with parent is bridge. Also $v$ know value of $g(u)$ for any child node $u$, hence $v$ can find out if bound $(u, v)$ is bridge or not.

As length of each $s'(v)$ at most $D + 1$, delay before getting first bit of $s'(u)$ from child node is at most $D - h(v)$, therefore the last stage completes in $O(D)$ rounds. The proof is complete.

**Theorem 3.** *Information about all bridges can be distributed in linear time.*

Let define *certificate of bridges* in similar way as certificate via BFS, but each symbol "1" that corresponds to a vertex which have bridge-bound with its parent we replace by symbol "2".

Easy to see that *certificate of bridges* fully describe all bridges in network. Therefore for complete the proof it is sufficient to build and distribute this certificate. And it is easy to make with simple modification of protocol from theorem 1.

As result, we get linear upper bound on time of combined protocol, that starts in network with a leader, builds tree-numeration (theorem 1), computes bridges (theorem 2), builds and distributes bridges-certificate.

Upper bound on summary traffic (number of non-empty messages what were send through all bounds) is $O(V^2 + E \log V)$, because all messages except constant symbols through each edge (echo-protocols and second rule in height-searching) and edges-numeration (stage 2 in theorem 2) go through tree edges, linear time and linear number of tree edges give summand $O(V^2 + E)$. Stage 2 in theorem 2 obviously costs $O(E \log V)$.

## *References.*

[1]. Lynch N.A. Distributed Algorithsm. San Frahcisco, CA: Morgan Kaufman, 1997

[2] Wattenhofer R. Principles of Distributed Computing. Lecture material. Swiss Federal Institute of Technology (ETH), Zurich, Switzerland, 2014. Available at http://dcg.ethc.ch/lectures/fs14/podc

[3] Vyalyi, M.N. & Khuziev, I.M., Distributed communication complexity of spanning tree construction, Probl Inf Transm (2015) 51: 49. https://doi.org/10.1134/S0032946015010068

[4] Vyalyi, M.N. & Khuziev, I.M., Fast protocols for leader election and spanning tree construction in a distributed network, Probl Inf Transm (2017) 53: 183. https://doi.org/10.1134/S0032946017020065

[5] Itai A., Rodeh M. Symmetry Breaking in Distributed Netwoks // Inform. Comput. 1990 v. 88 №1. P. 60-87.

[6] David Pritchard, An Optimal Distributed Edge-Biconnectivity Algorithm arXiv:cs/0602013